\documentclass[fleqn,usenatbib]{mnras}

\usepackage{newtxtext,newtxmath}
\usepackage[T1]{fontenc}
\usepackage{graphicx}
\usepackage{amsmath}
\usepackage{threeparttable}
\usepackage{pdflscape}

\newcommand{\lya}{Ly$\,\alpha$}
\newcommand{\lyb}{Ly$\,\beta$}
\newcommand{\lyg}{Ly$\,\gamma$}
\newcommand{\z}{$z$}
\newcommand{\msun}{\ensuremath{{\rm M}_\odot}}
\newcommand{\kms}{km s$^{-1}$}

\newcommand{\CIV}{{\rm C}$\,${\small\rm IV}\relax}
\newcommand{\OII}{{\rm O}$\,${\small\rm II}\relax}
\newcommand{\OVI}{{\rm O}$\,${\small\rm VI}\relax}
\newcommand{\SiIV}{{\rm Si}$\,${\small\rm IV}\relax}
\newcommand{\HI}{{\rm H}$\,${\small\rm I}\relax}
\newcommand{\NHI}{$N_{\rm H\,I}$}
\newcommand{\logNHI}{$\log N_{\rm H\,I}$}
 
\newcommand{\MHI}{$M_{\rm H\,I}$}
\newcommand{\logNCIV}{$\log N_{\rm C\,IV}$}

\newcommand{\hst}{{\em HST}}
\newcommand{\jwst}{{\em JWST}}

\title[Sunburst Arc Tomography]{Intergalactic Medium Tomography with the Sunburst Arc}

\author[M. A. Berg et al.]{
Michelle A. Berg,$^{1,2}$\thanks{E-mail: m.a.berg@tcu.edu}
John Chisholm,$^{2}$
J. Xavier Prochaska,$^{3,4,5}$
T. Emil Rivera-Thorsen,$^{6}$
\newauthor
Michael D. Gladders,$^{7,8}$
Keren Sharon,$^{9}$
Claus Leitherer,$^{10}$
J. J. Eldridge,$^{11}$
Matthew Bayliss,$^{12}$
\newauthor
Haakon Dahle,$^{13}$
Jane R. Rigby,$^{14}$
and Anne Verhamme$^{15,16}$
\\
$^{1}$Department of Physics \& Astronomy, Texas Christian University, Fort Worth, TX 76109, USA\\
$^{2}$Department of Astronomy, University of Texas at Austin, Austin, TX 78712, USA\\
$^{3}$Department of Astronomy and Astrophysics, University of California, Santa Cruz, CA 95064, USA\\
$^{4}$Kavli Institute for the Physics and Mathematics of the Universe (WIP), 5-1-5 Kashiwanoha, Kashiwa, 277-8583, Japan\\
$^{5}$Division of Science, National Astronomical Observatory of Japan, 2-21-1 Osawa, Mitaka, Tokyo 181-8588, Japan\\
$^{6}$The Oskar Klein Centre, Department of Astronomy, Stockholm University, AlbaNova, SE-10691 Stockholm, Sweden\\
$^{7}$Department of Astronomy \& Astrophysics, University of Chicago, 5640 South Ellis Avenue, Chicago, IL 60637, USA\\
$^{8}$Kavli Institute for Cosmological Physics, University of Chicago, 5640 South Ellis Avenue, Chicago, IL 60637, USA\\
$^{9}$Department of Astronomy, University of Michigan, 500 Church Street, Ann Arbor, MI 48109, USA\\
$^{10}$Space Telescope Science Institute,3 3700 San Martin Drive, Baltimore, MD 21218, USA\\
$^{11}$Department of Physics, University of Auckland, Private Bag 92019, Auckland, New Zealand\\
$^{12}$Department of Physics, University of Cincinnati, Cincinnati, OH 45221, USA\\
$^{13}$Institute of Theoretical Astrophysics, University of Oslo, P.O. Box 1029, Blindern, NO-0315 Oslo, Norway\\
$^{14}$Goddard Space Flight Center, 8800 Greenbelt Road, Greenbelt, MD 20771, USA\\
$^{15}$Univ Lyon, Univ Lyon1, Ens de Lyon, CNRS, Centre de Recherche Astrophysique de Lyon UMR5574, F-69230 Saint-Genis-Laval, France\\
$^{16}$Observatoire de Gen$\grave{e}$ve, Universit$\acute{e}$ de Gen$\grave{e}$ve, 51 Ch. des Maillettes, CH-1290 Versoix, Switzerland
}

\pubyear{2025}

\begin{document}
\label{firstpage}
\pagerange{\pageref{firstpage}--\pageref{lastpage}}
\maketitle

\begin{abstract} 

Gravitational lensing has transformed the field of gas tomography in the intergalactic medium (IGM) and circumgalactic medium (CGM). Here we use the brightest lensed galaxy identified to date, the Sunburst Arc (\z$\approx$2.37), to constrain the physical size of foreground absorbers at \z$\approx$2 in 2D. This galaxy is a confirmed Lyman continuum leaker, where its single leaking region is imaged 12 times over four separate arcs. The separations between the arcs allows for large scale tomography, while the distances between the images along an arc allow for small scale tomography. Using \hst/WFC3 UVIS G280 grism observations, we extracted the spectra of the leaking region and fit for absorbers detected along these lines of sight using a binary population and spectral synthesis (BPASS) model for the galaxy. We identified two partial Lyman limit systems (pLLSs) and one Lyman limit system (LLS) across the different spectra and measured their physical sizes. We find consistent \HI\ column densities across $\lesssim$2 kpc and an average \HI\ mass of $\approx$10$^3$ \msun\ for the absorbers. Given the strong \CIV\ lines associated with two of the absorbers, they are likely located within the CGM of foreground galaxies. The third absorber has no associated metal lines, so it is most likely within the IGM. This study provides the first tomography measurements of pLLSs/LLSs in the CGM and IGM at \z$\approx$2.

\end{abstract}


\section{Introduction} \label{sec:intro}

Galaxies constantly have gas flowing in, out, and around them, and their evolution depends on the availability of this material to continue star formation. Large-scale gas flows are found in the circumgalactic medium (CGM or gas in the halos of galaxies, \citealt{tpw2017}). New material can enter a galaxy halo from the intergalactic medium (IGM or gas in between large-scale structure in the universe), and gas can be ejected from a galaxy into the CGM or IGM through stellar outflows or active galactic nucleus outflows. 

Many observational studies have focused on constraining the mass of total gas available in the CGM and IGM at varying wavelengths (e.g., \citealt{danforth2008,werk2014,lehner2015,zheng2015,das2019,bregman2022,nicastro2023,khrykin2024}) as a way to understand how galactic outflows and inflows affect galaxy evolution. For instance, the CGM around an $L^*$ (Milky Way-like) galaxy holds $\gtrsim$$10^{10}$ \msun\ of gas within the virial radius (\citealt{werk2014,tpw2017}, and references therein), comparable to the stellar mass of this type of galaxy. However, poorly constrained assumptions are frequently made to obtain these estimates, such as the gas density and cloud structure, and photoionization modeling is usually necessary for a mass calculation (e.g., \citealt{prochaska2017,bregman2018}). When considering the total baryonic mass in the universe, galaxies house $\approx$7\% of the baryons, while the CGM and IGM contain $\approx$5\% and $\approx$53\%, respectively \citep{shull2012}. Within these estimates, there is still substantial uncertainty in the mass of gas in each temperature phase of the CGM and IGM \citep[e.g.][]{khrykin2024}. If we could define the typical size scales of hydrogen gas within these regions, this would increase the accuracy of the current baryon census.

From a theoretical standpoint, the results from observational studies have been puzzling. Cool ($T\sim10^4$ K) gas has a high covering factor (e.g., \citealt{bergeron1986,churchill2000,churchill2013,chen2010,kacprzak2013,nielsen2013,werk2013,dutta2020,huang2021}) and low volume filling factor (e.g., \citealt{stocke2013,cantalupo2014,werk2014,hennawi2015,stern2016,faerman2023}) in the CGM, yet current galaxy evolution theory asserts a ubiquitous hot ($T\sim10^6$ K) phase in the halo of massive galaxies (e.g., \citealt{maller2004,dekel2006,keres2009}). Somehow, these two phases coexist in the halos of galaxies and are also seen in the IGM \citep[e.g.][]{kovacs2019,manuwal2021}. 

\citet{mccourt2018} put forth a new theory on cloud sizes to reconcile theory and observation. As the hot, thermally unstable phase cools, it fragments into a cool, fine mist with cloudlets $<$0.1 pc in size. This mist fulfills the low volume filling factor and high covering factor as it is spread throughout the halo. Recent simulation efforts have tried to simulate a more realistic CGM and IGM through mass or spatial refinement methods \citep{peeples2019,vandevoort2019,nelson2020,mandelker2021,hummels2024,ramesh2024}, which has resulted in more small, cool cloud structures. Though sub-kpc resolution has been achieved, the simulations have not yet converged. To constrain cloud sizes and provide constraints for simulations, multiple measurements of the same gas cloud are necessary.

To study the IGM and CGM, a background QSO is often used to illuminate the foreground gas. These are pencil-beam observations, and usually only one background QSO pierces through a foreground galaxy halo. Therefore, statistical samples with many QSO sightlines at all different distances and angles around similar types of galaxies are used to understand the CGM and its properties (e.g., \citealt{steidel1992,bowen1995,tumlinson2011,bordoloi2014,tejos2014,borthakur2016,burchett2016,burchett2019,berg2019,berg2023,bielby2020,lofthouse2020,bouche2025}). Adapting the tomography technique used in \lya\ IGM studies (e.g., \citealt{dodorico2006,krolewski2017,mukae2020,giri2025}), other background sources (such as \citealt{lee2014}) have recently been used to perform tomography studies to reveal the extent and shape of gas distributions that focus on single galaxies or larger sections of a galaxy environment. 

Lensed QSOs offer two to four counter images (sightlines) around a single galaxy lens at small radii or multiple sightlines to investigate absorbing gas structures. Close QSO pairs with small angular separations have been used to determine the clumpiness of single absorbing systems (e.g, \citealt{hennawi2006,martin2010,rubin2015,bish2021,tuli2025}). Covering fractions of different ions and large-scale clustering (up to $\sim$ 900 kpc) have also been measured with these pairs (e.g., \citealt{rauch2001,ellison2004,lopez2007,chen2014,zahedy2016,zahedy2017,rubin2018,kulkarni2019,rudie2019,augustin2021}). Observations of this type constrain the coherence length of the gas. Finally, gravitational arcs have been used to study large swathes of the CGM or galaxy environments. Due to the technological advances of integral field unit (IFU) instrumentation (e.g., \citealt{bacon2006,bacon2010,morrissey2018}), the technique of arc tomography can continuously probe absorbing gas across hundreds of kpc (e.g., \citealt{lopez2018,lopez2020,mortensen2021,tejos2021,bordoloi2022,fernandez2022,afruni2023,lopez2024,berg2025}). 

The brightest gravitational arc found to date is the Sunburst Arc (also known as PSZ1 G311.65-18.48). It was discovered by \citet{dahle2016} and is at \z=2.3686 $\pm$ 0.0006, while the foreground lensing galaxy cluster is at \z=0.44316 $\pm$ 0.00035. The entire arc spans 55\arcsec\ and is composed of four separate arc segments. Along the segments, there are 12 bright knots. Further observations of one of these knots by \citet{rivera-thorsen2017} found that the \lya\ profile was triple-peaked, indicating a strong possibility of Lyman continuum (LyC) leakage. This \lya\ profile constrains the geometry of the surrounding medium of this star cluster to be an optically thick neutral medium with ionized channels for photons to escape (e.g., \citealt{zackrisson2013,behrens2014}). 

To confirm LyC leakage, \citet{rivera-thorsen2019} observed the Sunburst Arc with the {\it Hubble Space Telescope} (\hst) Advanced Camera for Surveys (ACS) in the F814W filter covering the non-ionizing stellar continuum (redward of 912\AA) and the F275W filter covering the ionizing stellar continuum (blueward of 912\AA). In the ionizing stellar continuum image, flux can still be seen coming solely from the 12 bright knots, confirming LyC leakage. Assuming the knots are images of the same source, they also conjectured the existence of foreground IGM absorption due to the varying escape fractions from each knot. Given the small separations between the knots, the authors determined that foreground gas could be probed on scales of $\sim$300 pc -- 3 kpc. 

Since these works, the Sunburst Arc has been extensively studied with lensing models \citep{pignataro2021,diego2022,sharon2022}, simple stellar population (SSP) models \citep{chisholm2019}, and spectral energy distribution models \citep{solimano2021}. The lensing models identify the leaking knots as images of the same star cluster within the galaxy. Many authors have studied the star cluster properties, such as the presence of stellar outflows \citep{vanzella2020,mainali2022,choe2025}, the potential for Wolf-Rayet stars and heavy nitrogen enrichment \citep{pascale2023,rivera-thorsen2024}, \lya\ escape \citep{owens2024,solhaug2025}, chemical abundances \citep{kim2023,welch2025}, and chemical evolution \citep{tapia2024}. 

The Sunburst Arc presents a unique opportunity to use another illuminating source to study foreground gas: the ionizing stellar continuum of stars that are producing LyC photons. In this paper, we present the Sunburst Arc spectra of the leaking region taken with \hst\ Wide Field Camera 3 (WFC3) UVIS slitless grism spectroscopy and constrain the size scale of foreground IGM and CGM absorbers through the use of SSP models to measure the \HI\ column density. Since this star cluster is imaged 12 times over four arcs, we can probe any foreground gas on pc to kpc scales depending on the redshift. Additionally, the geometry of the arcs allows for length and width estimates of these gas clouds. Due to the instrument resolution constraints, we cannot use the traditional arc tomography technique. We are also limited to investigating clouds with \logNHI\ $\ge$ 16.0 [cm$^{-2}$]. These types of clouds (also known as absorbers) are referred to as partial Lyman limit systems (pLLSs, 16.0 $\le$ \logNHI\ $<$ 17.2), Lyman limit systems (LLSs, 17.2 $\le$ \logNHI\ $<$ 19.0), super Lyman limit systems (SLLSs, 19.0 $\le$ \logNHI\ $<$ 20.3), and damped \lya\ absorbers (DLAs, \logNHI\ $\ge$ 20.3). 

This paper is organized as follows. Section~\ref{sec:obs} describes the \hst\ observations, and section~\ref{sec:redux} summarizes our data reduction process. In section~\ref{sec:modfits}, we present the BPASS models we used to measure the foreground absorber HI column densities. The spectra and absorber characteristics are detailed in section~\ref{sec:results}. In section~\ref{sec:discuss}, we compare our results to other surveys and discuss the implications of the measured length scales. Our main conclusions are listed in section~\ref{sec:concl}.

Throughout this paper, we adopt the \citet{planck2016} cosmology, notably $H_0$ = 67.7 \kms\ Mpc$^{-1}$, $\Omega$$_m$(\z\ = 0) = 0.309, and $\Omega$$_{\Lambda}$(\z\ = 0) = 0.691. All distances are in proper kpc. We refer to each image of the leaking region as a knot.

\section{Observations} \label{sec:obs}

The observations of the Sunburst Arc come from the \hst/WFC3 UVIS grism program PID 15966 (PI: Chisholm) taken in June 2021. The UVIS G280 grism in order one provides wavelength coverage from 1900-4500 \AA\ with a spectral resolution of $R = 70$ at 3000 \AA\ (dispersion of 13.0 \AA/pixel). The wavelength range extends to 1\micron, but the higher orders start to overlap at 4500 \AA. The field of view is 162\arcsec\ $\times$ 162\arcsec, and the pixel scale is 0\farcs04. 

\begin{figure*}
   \centering
    \includegraphics[width=\textwidth]{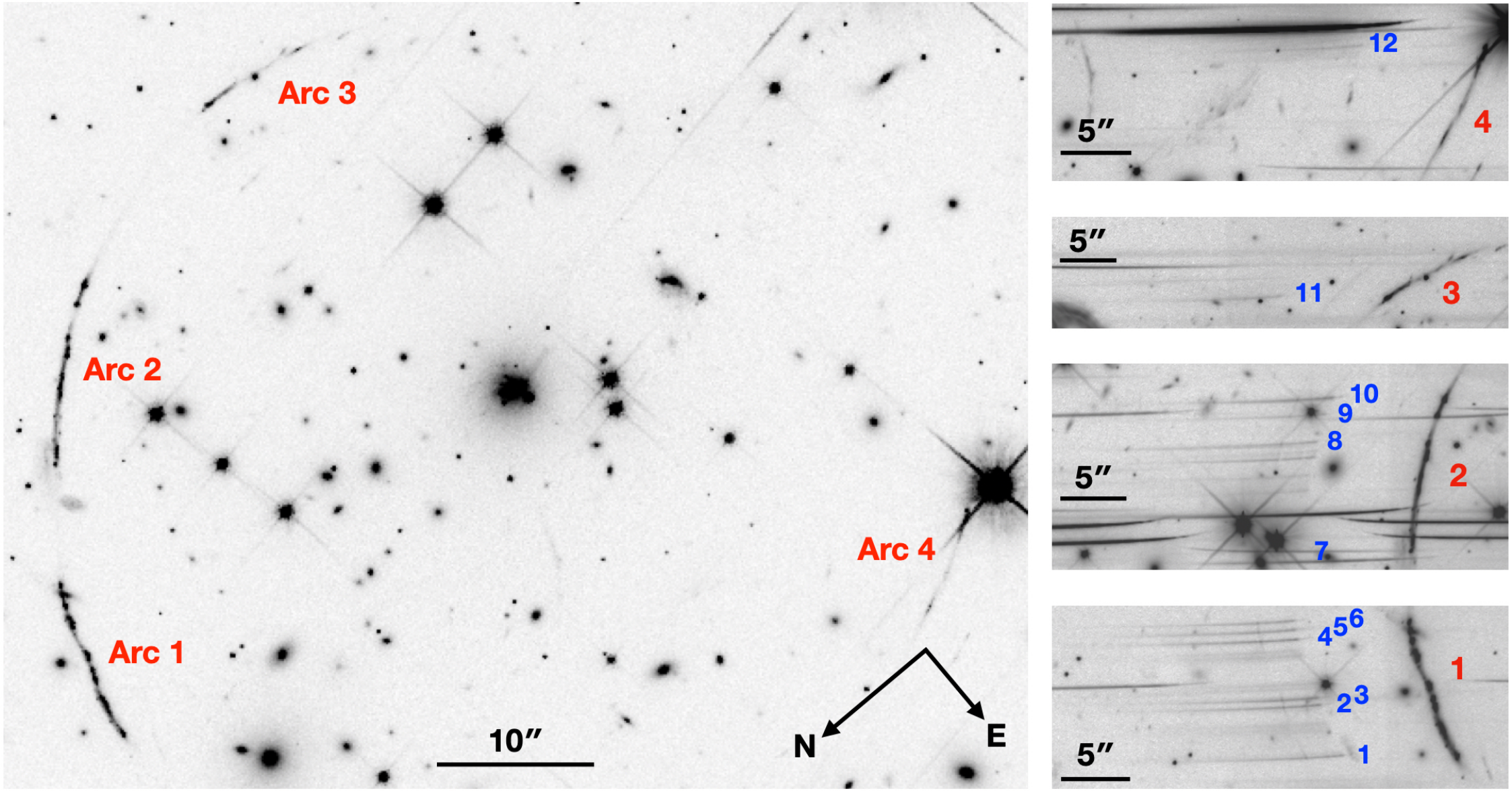}
    \caption{\hst/ACS F814W image of the Sunburst Arc (left, \citealt{rivera-thorsen2019}) and \hst/WFC3 UVIS G280 Beam A 2D spectra of each arc (labelled in red) with the 12 knot traces labelled in blue (right). Knot 7 is not detectable due to several bright foreground stars. Each knot is an image of the same leaking region in the galaxy.}
    \label{fig:hst}
\end{figure*}

Direct imaging of the field was obtained using the F300X filter, and the object was centered on chip two of the detector. The position angle of the observations was chosen so there was minimal contamination of arcs one and two (where 10/12 knots are located). A box dither was used in between visits. During visit three, there was a guide star failure causing two pairs of images and grism observations to be unusable. These exposures were reobserved in June 2022. Altogether, there are 12 F300X images (806 seconds total integration) and 27 G280 grism observations (83806 seconds total integration) of the field. In Fig.~\ref{fig:hst}, we show the \hst/ACS F814W image of the Sunburst Arc (PID: 15101, PI: Dahle) and the UVIS G280 grism Beam A observations of the leaking region for each arc.

\section{Data Reduction} \label{sec:redux}

We retrieved the {\it flc} calibrated direct and grism files from the Mikulski Archive for Space Telescope (MAST), where they were processed through the {\small CALWF3} data processing pipeline version 3.6.2. This pipeline preforms a bias subtraction, dark image subtraction, flat fielding correction, and charge transfer efficiency (CTE) correction to the raw exposure files. The files were retrieved after May 2021, so additional CTE and dark image corrections are already accounted for in the data \citep{prichard2022}. 

We utilized the code {\small GRIZLI}\footnote{https://github.com/gbrammer/grizli} (\citealt{brammer2019}, detailed in \citealt{abramson2020}) to perform a cosmic ray cleaning of the grism exposures and create drizzled, aligned, and combined direct and grism 2D images. {\small GRIZLI} is written for the IR side of the WFC3 detector, so we modified the code to also reduce the UVIS G280 files. On the right side of Fig.~\ref{fig:hst}, we show the drizzled G280 Beam A observation of the arcs. 

Since the {\small GRIZLI} code is not optimized for the spectral extraction of G280 traces, we created a modified version of the {\small XIDL} G280 reduction code\footnote{https://github.com/profxj/xidl/tree/master/HST/WFC3/G280} described in \citet{omeara2011} and \citet{lusso2018}. This code converts counts to flux and determines the wavelength solution of the spectral trace for the {\small GRIZLI} combined 2D images. There were several modifications we made to the extraction code. Though the code was identifying the correct source for extraction, the trace was being incorrectly placed above the source spectrum on the image. We added a y-pixel offset (average value of $-$120) to the trace solution to correct for this shift. There was also a shift in the wavelength solution. Previous spectral observations of the Sunburst Arc exist; we utilized the redshift of $z= 2.3686 \pm 0.0006$ reported in \citet{dahle2016} to center the \lya\ emission and determine the x-pixel offset (average value of $-$8) necessary in the extractions. As shown in \citet{rivera-thorsen2017}, the Sunburst Arc exhibits triple-peaked \lya\ with the central peak at the systemic velocity of the galaxy. Given the low resolution of this data, all three peaks are merged into one. The center of this line will still be at the systemic velocity.

The sky subtraction algorithm also needed to be altered due to the heavily crowded field. A 20-pixel window above and/or below the trace is designated as the sky subtraction region. Due to the close proximity of knots 2 and 3 and knots 4, 5, and 6, and the presences of diffuse stellar light near these traces, some of these sky regions cannot be used. Therefore, we added in the option of using the sky region from another extraction that aligns closely in x-pixel space. We used the sky region from knot 2 for knot 3 and the sky region from knot 6 for knots 4 and 5 (see Fig.~\ref{fig:hst}). In a few instances, we also increased the sigma clipping value in the sky region by 0.5 or 1.0 (default of 2.5$\sigma$) to more thoroughly remove bright stars.

Though there are updated UVIS trace calibration and sensitivity function files from 2021, we used the values from 2016.\footnote{https://www.stsci.edu/hst/instrumentation/wfc3/documentation/grism-resources/wfc3-g280-calibrations} The new files use different equations to parameterize the trace \citep{pirzkal2020}; implementing this into the code will be future work. This was not necessary for the current work because the 2016 calibration was able to cover the entire trace and only needed a small modification to accurately center the wavelength solution ($\approx$8 pixels on average). We extracted only Beam A (+1 order, left spectrum) due to the lower transmission of Beam C (-1 order, right spectrum) and the difficulty the code had in identifying the center of the Beam C traces. The {\small XIDL} code is optimized for QSO spectral extraction and normally runs with an optimal extraction. However, the knot traces are faint blueward of the Lyman continuum break wavelength of the Sunburst Arc. To make sure we were capturing the flux in this region accurately, we performed boxcar extractions (10 pixels in height) of the traces. During the extraction, we also masked regions of contamination on the traces such as foreground stars, galaxies, and diffuse light from other object traces.

After extracting and coadding the individual spectra for each knot from each {\it flc} file, we corrected for Milky Way extinction in the observed frame of the galaxy using {\small PYNEB}. We used the \citet{cardelli1989} reddening law with $R_V$ = 3.1, and $E(B-V)$ = 0.083, as reported in \citet{dahle2016}. The spectrum of knot 7 is unfortunately lost under two bright foreground stars. 

\section{Stellar Population Models and Fits} \label{sec:modfits}

\subsection{Stellar Models}\label{sec:models}

In this section, we utilize SSP models to constrain the absorber HI column densities and subsequent optical depths to identify consistent IGM structures within the knot spectra. For these fits, we utilize a high-resolution BPASS model v2.2.1 \citep{eldridge2017,stanway2018}. This model was fit to the Magellan/MagE \citep{marshall2008} observations of the Sunburst Arc from the MEGaSaURA Survey (\citealt{rigby2018} and Rigby et al., in prep.). This fit was made to the non-ionizing stellar continuum using restframe wavelengths 1240--1900\AA\ \citep{chisholm2019}.

The details of the fitting procedure are given in \citet{chisholm2019}, but here we summarize the steps. We assumed that the observed stellar continuum can be reproduced as a linear combination of $i$-numbered single age, single metallicity stellar population models ($M_i(\lambda)$) that were attenuated by a single foreground dust screen. We parameterized the dust screen with a single color-excess, $E(B-V)$, and the attenuation law, $k(\lambda)$, from \citet{reddy2016a}. The relative strength of each stellar population is set by a linear coefficient ($X_i$) multiplied by $M_i(\lambda)$. We solve for the $X_i$ and $E(B-V)$ values that best match the observed stellar continuum as:
\begin{equation}
F_{\text{obs}}(\lambda) = 10^{-0.4E(B-V)k(\lambda)} \times \Sigma_{i}X_{i}M_{i}(\lambda).
\end{equation} 
We used 50 individual $M_i(\lambda)$, with ages ranging from 1--40~Myr and metallicities from 0.05--2Z$_\odot$. The BPASS model was tested and accurately matched the observed stellar continuum features \citep[see figures 5, 6, 7 in ][]{chisholm2019}. For knot 5, \citet{chisholm2019} found a light-weighted average age of 3.6$\pm0.1$ Myr with a stellar metallicity of 0.66$\pm$0.05~Z$_\odot$ and an $E(B-V) = 0.15$. 

Since we do not have a high-resolution restframe UV spectrum of every single LyC emitting knot, we make the assumption that the stellar population properties for knot 5 are the same for all knots. The lens model suggests that each LyC emitting region is the same physical region within the galaxy, and the stellar populations should be nearly identical for each knot \citep{rivera-thorsen2019}. This was tested in \citet{mainali2022}, where five LyC regions were fit with the identical stellar population method described above. The individual stellar population properties were found to be consistent, within the uncertainties, for all five of the LyC emitting knots that were tested. This suggests that the same intrinsic stellar populations are measured in all LyC knots. The $E(B-V)$ values do have a slight variation, which \citet{kim2023} interpreted as different magnifications probing different amounts of dust. This is seen in Fig.~\ref{fig:arcs} by slight differences in slopes for each of the LyC emitting knots. Therefore, on the individual knot-by-knot analysis described below, we fit for $E(B-V)$ in each knot using the best-fit stellar models from the MagE observations. This corrects the slight offset between the general model and the knot-by-knot variation of the individual non-ionizing continua.

\begin{figure*}
   \centering
    \includegraphics[width=\textwidth]{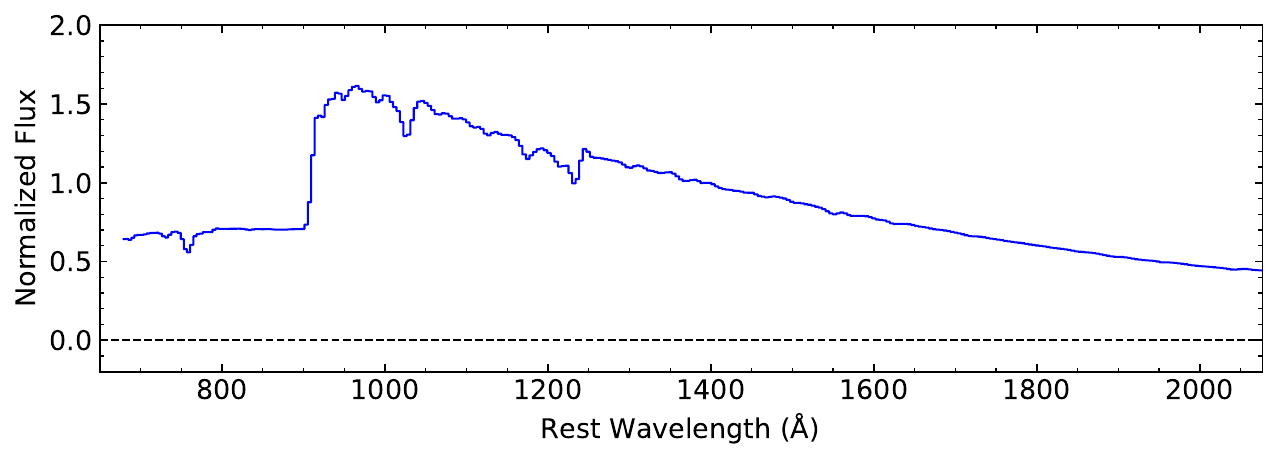}
    \caption{Convolved high-resolution BPASS model including the effect of dust attenuation used for absorber fits. We have interpolated the model onto the UVIS wavelength grid, convolved it with a Gaussian 1D filter with a FWHM of 10\AA, and corrected for Milky Way dust extinction. No emission lines are included in the model because it is only made from the star light.}
    \label{fig:models}
\end{figure*}

To make a fair comparison, we interpolated the model to the UVIS wavelength grid using the FluxConservingResampler in the {\small SPECUTILS} python package. The model was also convolved with a 1D Gaussian filter with a FWHM=10\AA\ to mimic the UVIS line spread function. Fig.~\ref{fig:models} displays the interpolated, convolved, general BPASS model with dust. The model spectrum is normalized to match the observations at wavelengths 1267-1276\AA. Milky Way dust is corrected for in both the observations and the model. We note the model does not include emission lines, which is why the \OVI\ $\lambda$$\lambda$1031.9,1037.6 emission peaks and \lya\ are not modeled. Following \citet{chisholm2022}, we calculate the $\beta$-slope for each spectrum using their equation 5 with the flux at 1300 and 1800 \AA. We then estimate the $E(B-V)$ value using our calculated $\beta$-slope and their equation 9 to create the individual knot models. Table~\ref{tab:ebv} presents the individual $E(B-V)$ values for each knot.

\begin{table}
\centering
\caption{Knot $E(B-V)$ Values}
\begin{threeparttable}
\begin{tabular}{lc}
\hline
\hline
Knot & $E(B-V)$ \\
\hline
1 & 0.22 $\pm$ 0.04\\
2 & 0.18 $\pm$ 0.04\\
3 & 0.12 $\pm$ 0.03\\
4 & 0.15 $\pm$ 0.04\\
5 & 0.22 $\pm$ 0.04\\
6 & 0.11 $\pm$ 0.06\\
8 & 0.32 $\pm$ 0.04\\
9 & $-$0.35 $\pm$ 0.17\\
10 & 0.11 $\pm$ 0.03\\
11 & 0.11 $\pm$ 0.07\\
12 & 0.24 $\pm$ 0.06\\
\hline
\end{tabular}
\begin{tablenotes}
\item Knot 9 is contaminated by a star, which causes this negative $E(B-V)$ value.
\begin{footnotesize}
\end{footnotesize}
\end{tablenotes}
\end{threeparttable}
\label{tab:ebv}
\end{table}

\subsection{Absorber Fits}\label{sec:absfits}

\begin{figure*}
   \centering
    \includegraphics[width=\textwidth]{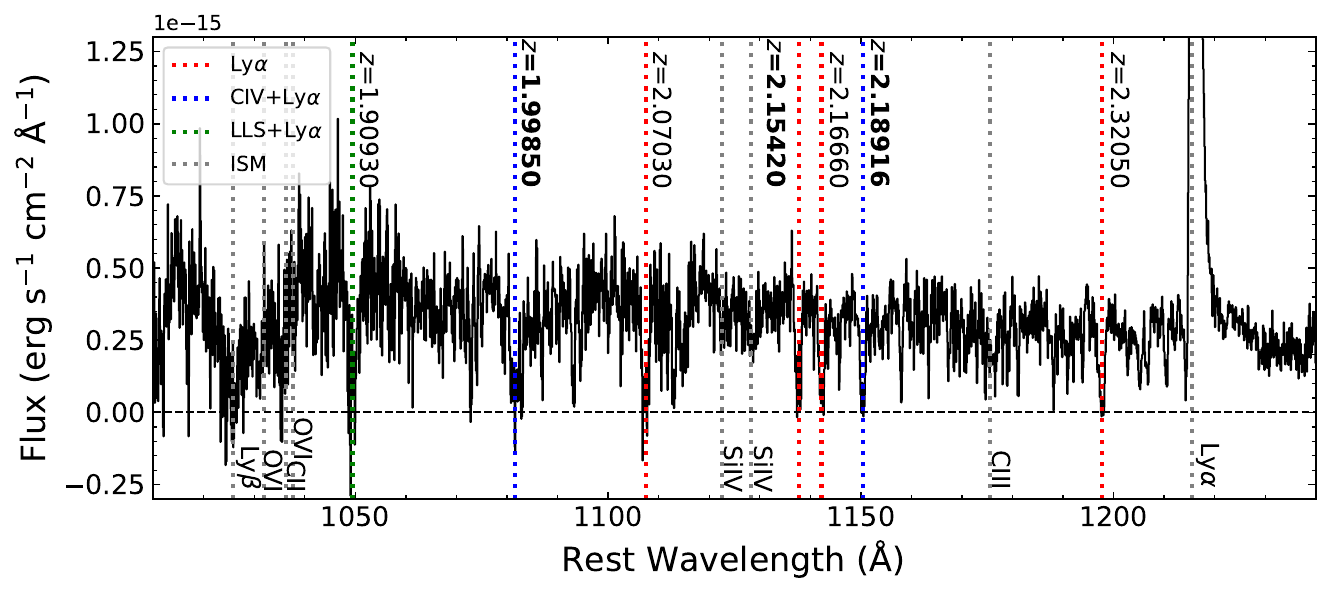}
    \caption{Magellan/MagE combined spectrum of the leaking region in the Sunburst Arc (knots 1, 4, 5, 8, 9, 10; Rigby et al. in prep). Strong galaxy interstellar medium (ISM) lines are shown in gray. Intervening foreground absorbers are marked by their redshift. All of the lines are either confirmed \lya\ associated with a \CIV\ absorber or LLS (blue and green dotted lines) or a potential \lya\ absorber (red dotted lines). We checked for these seven absorbers in each of the individual knot spectra; those in bold occurred repeatedly.}
    \label{fig:mage}
\end{figure*}

Seven doublet absorbers (\CIV\ or \SiIV) have been identified in the optical MagE spectra of knots 1, 4, 5, 8, 9, and 10 (Rigby et al. in prep). Additionally, there are five strong absorption lines blueward of the galaxy \lya\ line, which could be intervening \lya\ or a stronger \HI\ absorber (pLLS or higher \HI\ column density). Fig.~\ref{fig:mage} displays the combined MagE observations for the leaking images of the Sunburst Arc with all absorbers (confirmed and potential) labelled with their redshifts and the type of absorber. We checked for all of these systems in each knot when adding intervening absorbers into the models. However, several of the spectra have contamination in the region blueward of the galaxy restframe 912\AA, which limited our ability to fit for intervening absorbers. Table~\ref{tab:contams} summarizes these contamination issues, and the contaminants can be seen in the 2D spectra in Fig.~\ref{fig:hst}. 

\begin{table}
\centering
\caption{LyC Emitting Knot Traces With Contamination Blueward of 912\AA}
\begin{threeparttable}
\begin{tabular}{lcc}
\hline
\hline
Knot & Region & Contaminant \\
\hline
1 & 853$-$927\AA & Foreground galaxy\\
2 & 677$-$870\AA & Foreground star spikes\\
3 & 708$-$912\AA & Foreground star and spikes\\
8 & 770$-$814, 853$-$897\AA & Foreground star spikes\\
9 & 845$-$912\AA & Foreground star\\
\hline
\end{tabular}
\begin{tablenotes}
\item
\begin{footnotesize}
\end{footnotesize}
\end{tablenotes}
\end{threeparttable}
\label{tab:contams}
\end{table}

We started the fits by adding the break at the Lyman limit from the galaxy interstellar medium into the models. (We do not report the galaxy \logNHI\ values in this paper, but save this analysis for Berg et al. 2025b, in prep.) Fig.~\ref{fig:llsmodel} shows the general absorber model spectrum we used in these fits; this model is interpolated to the UVIS wavelength grid and convolved with a 1D Gaussian filter with a FWHM=10\AA\ to mimic the UVIS line spread function prior to being added to the BPASS model. Once the model adhered to the observations at 912\AA, we inspected the location of the Lyman break for each absorber. If there was a blueward flux decrease, we estimated the absorber column density and added it to the model. If an absorber did not improve the overall model fit to the data (e.g., it decreased the blueward flux too much to match the observations), we did not include it. To estimate the absorber column density at the absorber Lyman limit, we used the standard equation \NHI=$\tau_{\rm LL}/\sigma_{\rm LL}$, where $\tau_{\rm LL}$ is the optical depth at the Lyman limit, and $\sigma_{\rm LL}$ is the absorber cross-section of the hydrogen atom at the Lyman limit. At values below \logNHI=16, absorbers will not impose a decrease of flux in the spectrum. Therefore, we do not add any of these absorbers into the model. 

We do not expect perfect fits to the observations because very low resolution data smears and smooths out a lot of the features in a spectrum. The model fits presented here follow the general shape of the observed flux and at times are above or below the observations. We compared our fits to those in \citet{lusso2018} where they were fitting UVIS spectra of QSOs, and they also do not have complete agreement between their models and the observations. However, we stress that we are not overfitting the data. We are only including absorbers seen within the MagE data and only those absorbers that improve the overall model fit. Changes of 0.1 dex in the column density can be determined with this data. This was evaluated by eye though offsetting the model in different increments
to be inconsistent with the observations; 0.1 dex was the lowest increment that caused a noticeable change to the fits. 

\section{Results} \label{sec:results}

In this section, we present the reduced galaxy spectra for the 11 images of the leaking region and an example BPASS model fit. (The others are included in Appendix~\ref{sec:appa}.) We then calculate the absorber size scales and estimate the \HI\ mass. 

\subsection{Galaxy Knot Spectra}\label{sec:galspec}

\begin{figure*}
   \centering
    \includegraphics[width=\textwidth,height=8.55in]{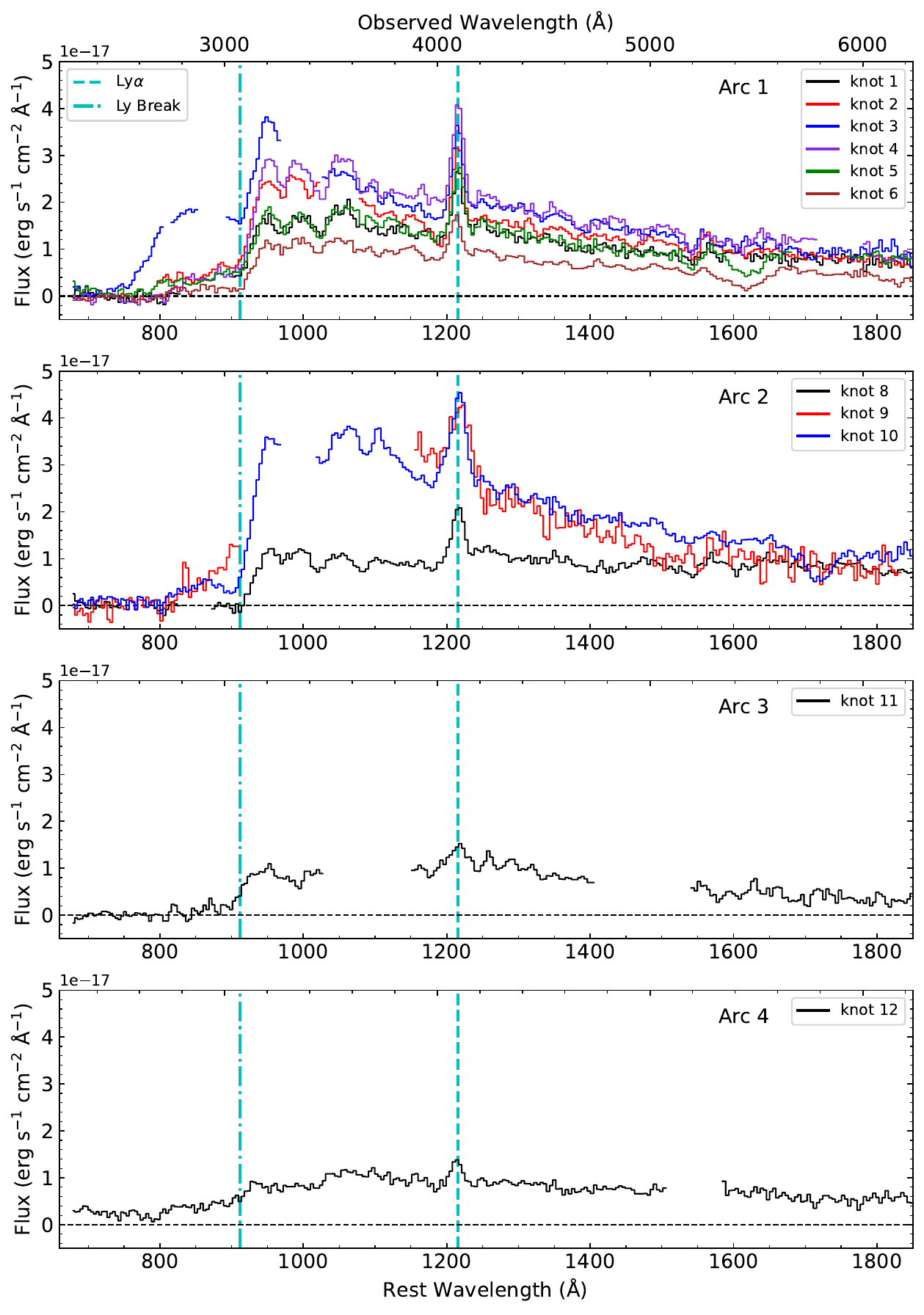}
    \caption{\hst/WFC3 UVIS G280 spectra of each knot for the different arcs at rest wavelengths. \lya, \lyb, \lyg, and the Lyman break are prominent features in these spectra. Regions that were masked due to contamination appear as breaks in the spectral flux. We do not plot the left-most end of the spectra due to an exponential increase in the sensitivity function; this increase in flux is not real. Many of the knots exhibit LyC flux, which can be used to probe foreground IGM gas.}
    \label{fig:arcs}
\end{figure*}

In Fig.~\ref{fig:arcs}, we present the 11 spectra of the extracted leaking region for each arc. (Knot 7 is unable to be identified and extracted in the observations due to bright foreground stars.) We only plot the spectra down to rest wavelength 677\AA\ because the UVIS sensitivity function grows exponentially positive to the left of this wavelength; this causes an artificial increase in the flux at the lowest wavelengths. Since the LyC emitting knots are not extremely bright, we are not concerned with overlapping higher-order spectra redward of rest wavelength 1335\AA. Regions of missing flux have been masked due to contamination of other objects in the field falling on the trace. Almost all of the knots show foreground absorption around 790\AA\ by an LLS (or higher column density absorber) at \z$\approx$1.99 that is removing all of the flux blueward of this wavelength. This confirms the hypothesized absorption in \citet{rivera-thorsen2019}. 

\subsection{Absorber Characteristics}\label{sec:abschar}

\begin{figure*}
   \centering
    \includegraphics[width=\textwidth]{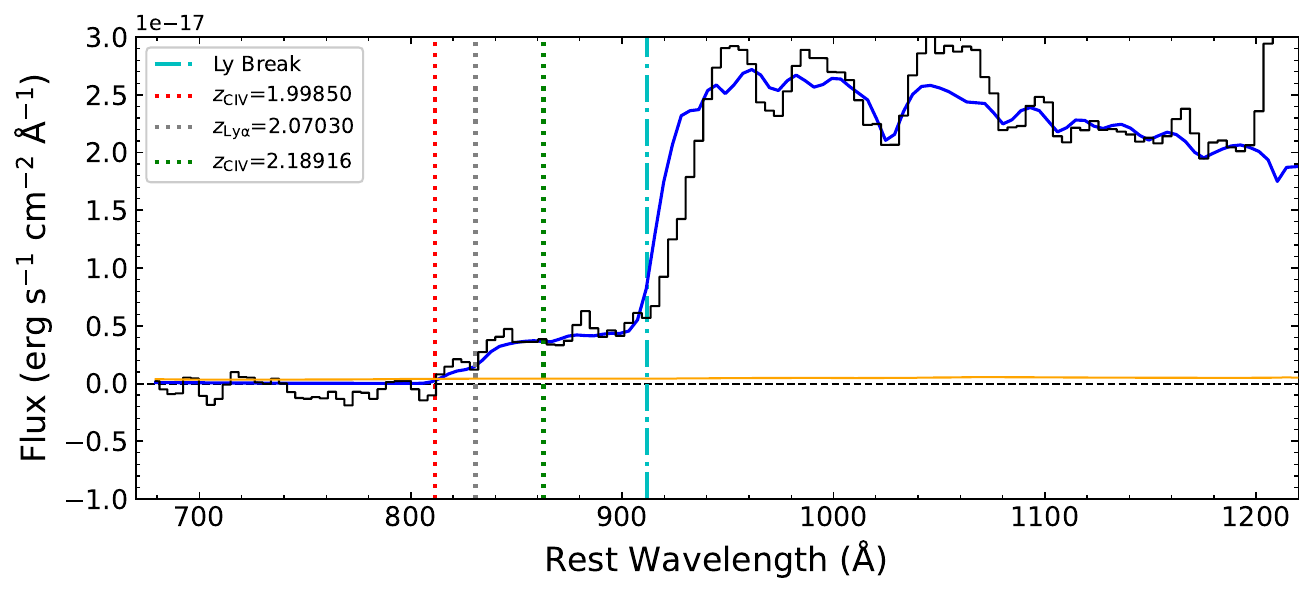}
    \caption{BPASS model fit to knot 4 spectrum. The galaxy Lyman break is denoted by the dashed and dotted line. The Lyman break of two intervening absorbers that appear in more than one knot spectrum are shown with the colored dotted lines. The Lyman break of a singly-occurring absorber is shown with the gray dotted line. We do not fit the model to the data, but rather only add to the model the absorbers that are identified in the MagE spectrum if they improve the fit.}
    \label{fig:modfit}
\end{figure*}

Fig.~\ref{fig:modfit} shows the model absorber fits for knot 4, and Table~\ref{tab:abshi} summarizes the foreground absorber \logNHI\ fits for each knot. All but one potential absorber shown in Fig.~\ref{fig:mage} was identified in at least one spectrum. The absorption at \z=2.32050 must either be along the line of sight of one of the knots with contamination or due to a different ion. There are only three strong, recurring absorbers (one LLS and two pLLSs) identified in the spectra with consistent column densities: two \CIV\ absorbers with associated \lya\ at \z=1.99850 and 2.18916 (identified in Rigby et al., in prep) and a \lya\ absorber at \z=2.15420. The \logNHI\ lower limits for the \CIV\ absorber with associated \lya\ at \z=1.99850 are due to the fact that the blueward flux from the galaxy decreases to zero and does not recover. We infer \logNHI\ $>$ 17.2, and therefore characterize it as an LLS.

\begin{figure}
    \centering
    \includegraphics[width=3.3in,height=1.3in]{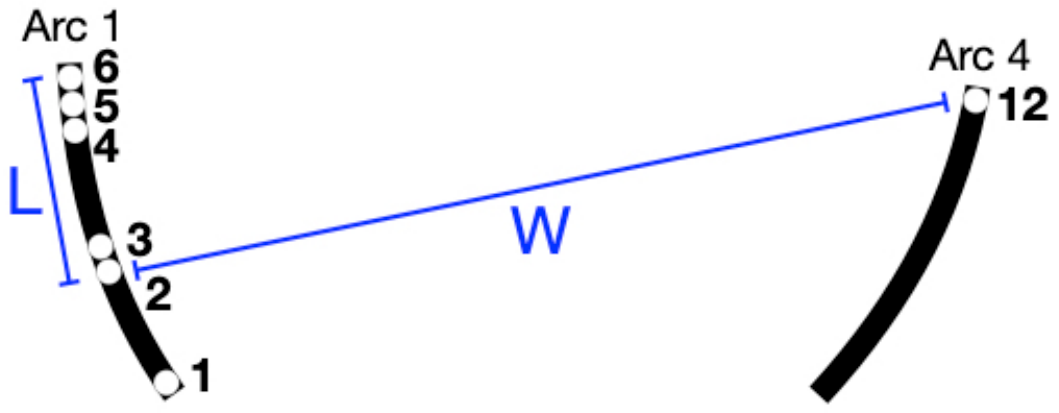}
    \caption{Cartoon showing the \z=2.15420 absorber tomography measurements. The LyC emitting knots are displayed on the arcs as white dots and labeled by number. Absorber lengths are measured along arcs 1, 2, and 3. Absorber widths are measured across the field of view between arcs 1, 2, 3 and arc 4. These are the first 2D size estimates of pLLSs/LLSs at \z$\approx$2.}
    \label{fig:tomo}
\end{figure}

\begin{figure}
   \centering
    \includegraphics[width=3.2in,height=2.5in]{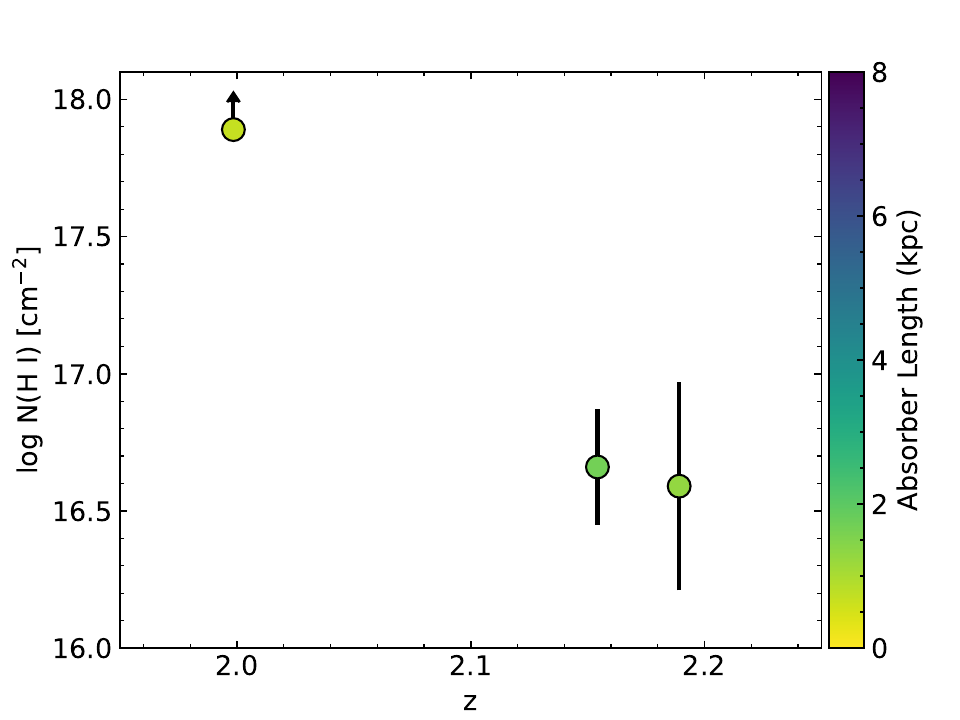}
    \caption{Average \HI\ column density versus redshift for the recurrent absorbers. The points are colored by the de-lensed length of the absorber. The absorber de-lensed lengths are also lower limits, as the absorbers could extend beyond where the knots probe the foreground gas. We are able to probe pLLS and LLS absorbers on kpc scales at \z$\approx$2 for the first time.}
    \label{fig:abslength}
\end{figure}

Since there is a unique lensing geometry for this system, we can leverage it to perform tomography of these pLLS and LLS foreground absorbers for the first time at \z$\approx$2. The multiple knots over multiple arcs give us the rare opportunity to simultaneously probe the small and large scale absorption properties of the gas. A cartoon of the tomography measurement locations is shown in Fig.~\ref{fig:tomo}. We estimate the absorber length scales by measuring the angular distance between the first and last knot where the absorber is seen (with consistent \logNHI\ within the error bars) and convert this to kpc at the redshift of the absorber. We then calculate the de-lensed length using equation S4 in \citet{rivera-thorsen2019}. For instance, the $z=2.18916$ absorber is detected in knot 2, 4, 5, but not in knot 6. This gives us a lower limit of the size as the distance between knot 2 and knot 5 of 1.3 kpc, since the knot 1 Lyman continuum region is contaminated. Utilizing Arc 4, we also estimate the width of the absorbers. This allows for a 2D size estimate instead of only 1D. For the $z=2.18916$ absorber, we measure between knot 2 (detection) and knot 12 (non-detection) for a upper limit of the width of 12.5 kpc. Table \ref{tab:abschar} provides estimates of the absorber length and width scales (lensed and de-lensed) at the redshift of the absorber, and Fig.~\ref{fig:abslength} displays the average absorber \logNHI\ with respect to redshift. There is consistency in the column density of the absorbers over hundreds of pc, with the absorbers measuring $\approx$1--2 kpc in length. We consider these lengths as lower limits and the widths as upper limits because we do not have enough sightlines to probe these absorbers at longer or shorter distances, respectively. To estimate the \HI\ mass of the absorbers, we use the equation \MHI=$m_{p}$\NHI$\pi$$r^2$, where we assume a circular geometry with half of the de-lensed length or width for the cloud radius. The masses are reported in Table~\ref{tab:abschar} and range from $\approx$$10^2-10^6$ \msun.

\begin{table*}
\centering
\caption{Absorber \HI\ Column Densities}
\begin{threeparttable}
\begin{tabular}{lcccccccc}
\hline
\hline
$z_{\rm abs}$ & Type & Knot 2 & Knot 4 & Knot 5 & Knot 6 & Knot 10 & Knot 11 & Knot 12 \\
\hline
2.32050 & \lya & \dots          & \dots          & \dots          & \dots          & \dots          & \dots          & \dots\\
2.18916 & \CIV+\lya & 16.57$\pm$0.16 & 16.38$\pm$0.31 & 16.74$\pm$0.14 & \dots          & \dots          & \dots          & \dots\\
2.16660 & \lya & \dots          & \dots          & \dots          & \dots          & \dots          & 17.20$\pm$0.23 & \dots\\
2.15420 & \lya & 16.71$\pm$0.16 & \dots          & 16.45$\pm$0.33 & 16.77$\pm$0.32 & \dots          & \dots          & 16.19$\pm$0.27\\
2.07030 & \lya & \dots          & 17.27$\pm$0.13 & \dots          & \dots          & \dots          & \dots          & \dots\\ 
1.99850 & \CIV+\lya & \dots          & $>$17.89       & \dots          & $>$17.68       & 17.29$\pm$0.21 & $>$17.75       & 16.48$\pm$0.19\\
1.90930 & LLS+\lya & Masked         & \dots          & 17.33$\pm$0.15 & \dots          & \dots          & \dots          & \dots\\  
\hline
\end{tabular}
\begin{tablenotes}
\item Entries with \dots\ signify an absorber could not be fit even though there is flux detected; see $\S~\ref{sec:absfits}$ for more information. Entries labeled "Masked" are regions where there is no flux due to contamination masking. We mark column densities as lower limits if the flux decreases to zero due to the absorber and the blueward observed flux does not recover. The potential absorber at \z=2.32050 is not detected in these knot spectra; it could be along the line of sight for the knots that have contaminated spectra. 
\begin{footnotesize}
\end{footnotesize}
\end{tablenotes}
\end{threeparttable}
\label{tab:abshi}
\end{table*}

\begin{table*}
\centering
\caption{Absorber Characteristics}
\begin{threeparttable}
\begin{tabular}{lccccccccccc}
\hline
\hline
$z_{\rm abs}$ & \logNHI & \logNCIV & Length & Length & De-lensed & Knots & Width & Width & De-lensed & Knots & \HI\ mass \\
 & [cm$^{-2}$] & [cm$^{-2}$] & (\arcsec) & (kpc) & (kpc) & & (\arcsec) & (kpc) & (kpc) & & \msun \\ 
\hline
2.18916 & 16.59$\pm$0.19 & $>$14.33 & 5.7 & 48.3 & 1.3 & 2$-$5 & 56.5 & 478.9 & 12.5 & 2$-$12 & 4.1$\times 10^2$--3.8$\times 10^4$\\
2.15420 & 16.66$\pm$0.29 & \dots & 6.3 & 53.8 & 1.7 & 2$-$6 & 56.5 & 480.1 & 15.2 & 2$-$12 & 8.3$\times 10^2$--6.6$\times 10^4$\\
1.99850 & $>$17.89 & $>$14.51 & 1.3 & 11.4 & 0.7 & 4$-$6 & 57.7 & 495.0 & 29.5 & 4$-$12 & 2.4$\times 10^3$--4.2$\times 10^6$\\
\hline
\end{tabular}
\begin{tablenotes}
\item The average \logNHI\ value is listed for each absorber. The absorber lengths are lower limits because we do not have another source to probe the absorber above or below the span of the currently detected knots. The absorber widths are all upper limits; we use whichever detected knot gives the most constraining width. The \HI\ masses listed are calculated assuming the de-lensed absorber length scale as the cloud diameter and then the width. The true mass is more likely to be within these limits.
\begin{footnotesize}
\end{footnotesize}
\end{tablenotes}
\end{threeparttable}
\label{tab:abschar}
\end{table*}

\section{Discussion} \label{sec:discuss}

Using \hst\ UV observations of the multiply-imaged LyC leaking region of the gravitationally-lensed Sunburst Arc, we identified three foreground pLLSs/LLSs at \z$\approx$2 in more than one spectrum. Two of these absorbers are \CIV\ absorbers with associated \lya, while the third is a \lya\ absorber. We performed 2D IGM tomography for the foreground absorbers and found the de-lensed lengths of the clouds are all $<$2 kpc. Assuming a spherical geometry, the absorbers have an average \HI\ mass of $\approx$$10^3$ \msun. 

\subsection{Absorber Properties}\label{sec:estimates}

\citet{rivera-thorsen2019} correctly hypothesized that the varying LyC flux they measured in each leaking region of the Sunburst Arc was due to foreground absorbers near \z$\approx$2, and that the multiple sightlines would allow this gas to be probed on hundreds of pc to a few kpc scales. Our results are the first tomography measurements of pLLSs/LLSs at \z$\approx$2. At this redshift, the Lyman limit is shifted out of the wavelength range of \hst\ and not yet shifted into the optical. Due to this observing constraint, there does not exist a large absorber survey for which we can compare our results at the same redshift. 

The KODIAQ-Z survey \citep{lehner2016,lehner2022} covered 2.2 $\lesssim$ \z\ $\lesssim$ 3.6 and studied the properties of over 200 absorbers with 14.6 $\le$ \logNHI\ $<$ 20. Using {\small CLOUDY} models \citep{ferland2013} to calculate the hydrogen density and total hydrogen column density, the absorber lengths scales can be estimated as $l = n_{\rm H}/N_{\rm H}$. Our three absorbers lie in the \logNHI\ ranges for pLLSs and LLSs, which have estimated lengths from the KODIAQ-Z survey of 4 $\lesssim l \lesssim$ 150 kpc and 18 $\lesssim l \lesssim$ 288 kpc, respectively. Absorber widths were not calculated in the survey. These ranges are larger than what we calculate for the foreground absorbers, and the authors note that the length scales show the largest dispersion out of all of the physical quantities they determined for the KODIAQ-Z survey. Though these scales are solely based on the modeling, it shows that this gas likely probes a variety of structures at \z$\approx$2--3. Our absorber lengths are lower limits, since we do not have uncontaminated sightlines on both sides of the absorbers to determine the full structure length. As such, the true values may be more in-line with those calculated in the KODIAQ-Z survey.

\begin{figure}
   \centering
    \includegraphics[width=3.5in,height=2.5in]{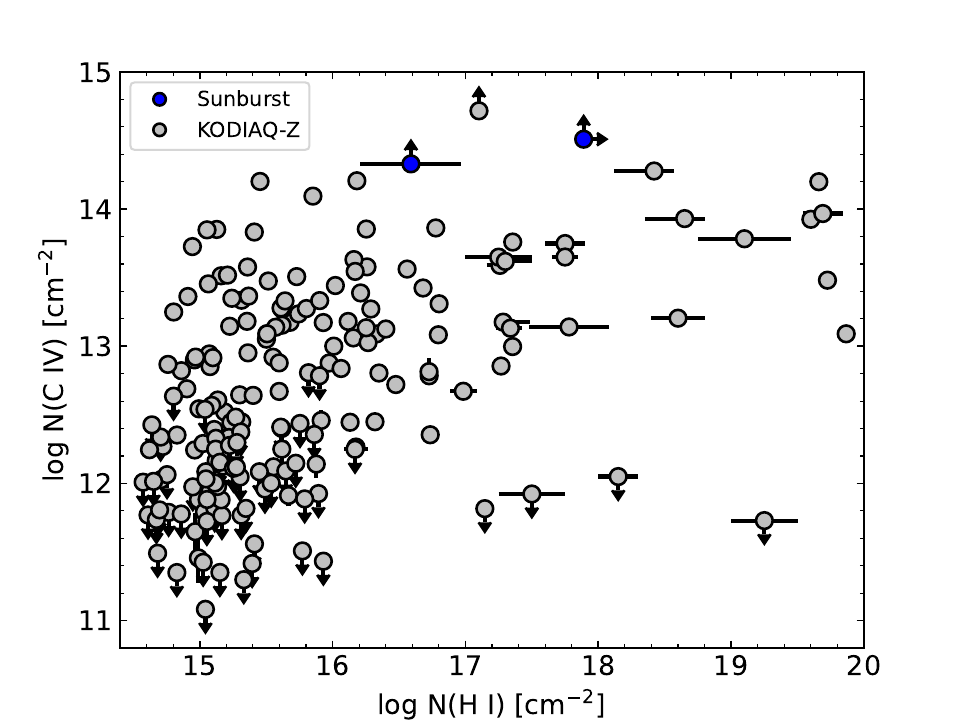}
    \caption{Absorber \CIV\ column density versus \HI\ column density. We compare the column densities of the two \CIV\ absorbers with associated \lya\ to those from the KODIAQ-Z sample \citep{lehner2022}, which lies at slightly higher redshifts (2.2$\lesssim\z\lesssim$3.6). The \CIV\ absorbers with associated \lya\ exhibit large metal column densities compared to the majority of the KODIAQ-Z sample. These \CIV\ absorbers with associated \lya\ are multiphase and therefore likely located within the CGM of a foreground galaxy halo instead of the IGM.}
    \label{fig:absciv}
\end{figure}

In Fig.~\ref{fig:absciv}, we compare the \CIV\ and \HI\ column densities of our two \CIV\ absorbers with associated \lya\ to those measured in the KODIAQ-Z survey. The \CIV\ absorbers with associated \lya\ have metal-line column densities well above the average value seen in the survey, with only one absorber residing at a higher value. Utilizing the ionization correction fractions for \HI\ and \CIV\ from \citet{simcoe2004}, we estimate the metallicity of the two \CIV\ absorbers with associated \lya. Both give super-solar values, values that are not found for absorbers in the KODIAQ-Z survey. These results point to gas that is collisionally-ionized and multiphase, instead of the usual photoionization assumption when estimating metallicities. 

Given how our absorber properties compare to the general population at similar redshift, we can now explore the locations of the gas. The \lya\ absorber has no known metal lines associated with it in the MagE spectrum, and is thus most likely residing in the IGM. We note this absorber could also potentially be inflowing gas in the CGM of a foreground galaxy (e.g., \citealt{fg2011,fumagalli2011}), but this cannot be confirmed (see \S~\ref{sec:hosts}). \CIV\ absorbers can reside in either the IGM (e.g., \citealt{manuwal2021}) or CGM (e.g., \citealt{burchett2015,burchett2016}). However, the metal column densities far exceed the KODIAQ-Z survey average. Such metal-enriched, multiphase gas must have been processed through a galaxy, placing the most likely location for these two absorbers in the CGM. 

\subsection{Column Density Variability}\label{sec:otherestimates}

When identifying recurrent absorbers within the knot spectra, we checked for consistent \logNHI\ values within the error bars from spectrum to spectrum. Even though knot 12 shows absorption at the same redshift as the \lya\ absorber and one of the \CIV\ absorbers with associated \lya, we do not think this is a continuation of the same absorber over tens of kpc. This is in large part due to how well the BPASS model fits to the observations. The slight absorption added improves the fit, but does not change it significantly. This implies there is little intervening gas along this sightline compared to the others. However, we are unable to fully rule this out, since we do not have any other sightlines to better constrain the cloud widths. 

For the $z=1.99850$ \CIV\ absorber with associated \lya, the \logNHI\ values were not consistent for every detection. It may be one large cloud that exhibits varying \HI\ density or two separate clouds. Knots 4, 6, and 11 have consistent \logNHI\ lower limit values, while knot 10 provides a detection with error bars. At the highest allowed value, knot 10 is $\approx$0.2 dex below the knot 6 value. The de-lensed distance between knots 6 and 10 is 10.3 kpc, implying a total absorber length of 18.1 kpc from knot 4 to 11. This length is similar to the cloud widths in Table~\ref{tab:abschar}, which are all upper limits due to the knot 12 non-detections. Unfortunately, we cannot use knots 7, 8, or 9 to measure the \logNHI\ in the region between knots 6 and 10 to determine which hypothesis is true due to contamination issues. The \CIV\ column densities cannot help either, as they are all lower limits (Rigby et al., in prep). Given the limited information we have, we choose the more conservative interpretation of two different clouds. 

Though we have used the assumption that consistent \logNHI\ values can identify recurring absorbers, the \HI\ density of these gas clouds is certainly not uniform as we have assumed here. Inflowing material into a galaxy halo is often seen in simulations in the form of large-scale streams with lengths on the order of the virial radius (e.g., \citealt{fg2011,fumagalli2011}). Outflowing material can entrain small cool clouds, and recycling can occur where the gas condenses and rains back down onto the galaxy. The \CIV\ absorbers with associated \lya\ are most likely in this latter category, where smaller size scales are expected. From our measurements, we see that on scales $\lesssim$2 kpc the cloud \HI\ column densities do not vary significantly. However, on scales $\gtrsim$10 kpc it is difficult to determine whether or not the same structure is being probed by the sightlines, while exhibiting varying density. 

Other surveys have seen similar results. \citet{bordoloi2022} observed two DLAs in the spectrum of a galaxy lens with Keck/KCWI \citep{morrissey2018} to determine how the ion column densities varied over the structures. The DLA at \z$\approx$2.5 exhibited varying \logNHI\ of $\approx$0.9 dex over 2-3 kpc (maximum extent of 17 kpc); the metal lines also varied over this length. The second DLA at \z$\approx$2.1 varied only by $\approx$0.25 dex in \logNHI\ over 6 kpc (maximum extent of 18 kpc). \citet{lopez2018} combined observations of VLT/MUSE \citep{bacon2006,bacon2010} and Magellan/MagE \citep{marshall2008} to study how the equivalent width (EW) of MgII varied using two segments of a background galaxy lens. The absorber at \z$\approx$0.98 had a varying EW of $\approx$1.1\AA\ over 29 kpc in the eastern section of the lens and $\approx$0.71\AA\ over 10 kpc in the western section. \citet{lopez2020} identified another MgII absorber using VLT/MUSE and Magellan/MagE observations of the Sunburst Arc. At \z$\approx$0.73, the EW varied by $\approx$1.94\AA\ over 23 kpc in Arc 2. No detections were measured in Arc 1. \citet{tejos2021} and \citet{mortensen2021} observed the same galaxy lens and \z$\approx$0.77 MgII absorber: Tejos et al. with VLT/MUSE and Magellan/MagE (1 psuedo-spaxel $\approx$ 1\farcs0 on the arcs and 0\farcs3 on the galaxy) and Mortensen et al. with Keck/KCWI (1 spaxel = 0\farcs7). The sightline has a north and south lens segment, with the absorber host galaxy in the middle. Tejos et al. reported the largest change in EW of $\approx$1.27\AA\ over 5 kpc in the northern segment, $\approx$1.26\AA\ over 5 kpc in the southern segment, and $\approx$1.61\AA\ over 6 kpc on the galaxy. Mortensen et al. found the largest change in EW of $\approx$0.8\AA\ over 2.3 kpc in the northern segment, $\approx$1.5\AA\ over 17 kpc in the southern segment, and consistent values within the error bars on the galaxy. The northern segment results are consistent with each other, but the southern segment results differ greatly. This is likely due to the different spatial resolutions between the two instruments. The KCWI observations cover 66 pixels, while the MUSE observations cover 21 pixels. 

Lensed QSOs have also been used to determine absorber length scales. \citet{zahedy2016} observed three lensed QSO systems at \z$\approx$0.4-0.7. The first had only upper limits for MgII EW (four images), the second measured a change of $\approx$0.63\AA\ over 23 kpc (two images), and the third measured $\approx$0.8\AA\ over 3.2 kpc (two images). In \citet{zahedy2017}, follow-up \HI\ observations were conducted for the third system; the column density was consistent between the two images, and the kinematics were also similar. \citet{rudie2019} also observed a lensed QSO system with two images at \z$\approx$2.4 separated by 400 pc in the Keck Baryonic Structure Survey (KBSS, \citealt{rudie2012}). The metal lines show variability in their line profiles for the different components, but the total velocity spread of the absorption is essentially identical. This implies the total gas structure is larger than 400 pc, though the components could be smaller cloudlet structures with scales less than 400 pc. \citet{augustin2021} used VLT/MUSE observations and simulations to interpret the varying MgII EWs seen in a quadruply-lensed QSO system. The \z$\approx$1.2 absorber showed coherence over at least 6 kpc, and the \z$\approx$1.4 absorber showed EW variations of $\approx$0.36\AA\ over $\approx$2 kpc. The interpretation is that the \z$\approx$1.2 absorber traces an inflowing gas filament, while the \z$\approx$1.4 absorber traces an outflowing clump. Though not a lensed galaxy, \citet{cooke2015} observed a foreground DLA at \z$\approx$2.4 in the spectrum of a background galaxy (\z$\approx$2.8) that is 40 kpc across with VLT/VIMOS \citep{lefevre2003}. The 2D spectrum shows complete absorption at \lya, but different sections of the galaxy cannot be compared with this long-slit data. 

Another approach for estimating gas length scales over larger distances is through paired galaxies. To study the small-scale structure of the IGM between 2$\lesssim$\z$\lesssim$4, \citet{rorai2017} utilized QSO pairs at similar redshifts. There are several pairs at \z$\approx$2 that are $\approx$40--145 kpc apart showing \lya\ absorption lines at the same redshifts. Even though the column densities are not always similar (i.e., showing varying densities), the sightlines are clearly probing the same large-scale gas structures. \citet{prusinksi2025} utilized galaxy pairs with separations of 8--250 kpc from the KBSS at \z$\approx$2 to study the extent of \lya\ and \CIV\ in the CGM. These ions were ubiquitously detected to 250 kpc. \citet{scarlata2025} studied the IGM \lya\ forest transmission at \z$\approx$2.5 in the COSMOS field with a sample of 268 background galaxies. They found the large-scale structures can be over several arcminutes to a degree in length (Mpc scales).  

In terms of \logNHI, there seems to be consistent measurements for absorbers with length scales below 2 kpc and large variations for length scales greater than 10 kpc. In between these distances, the measurements may or may not be consistent. The HI may instead be tracing the coherence length of total gas structures, not the actual individual cloudlet sizes, as conjectured by \citet{rudie2019}. This is consistent with our results. The metal ions, which varying widely over small and large length scales, seem to be tracing the cloudlet sizes better than the HI. However, there are few sub-kpc size scale estimates for the CGM due to instrumentation limitations and the rarity of multiple sightlines close together. The spatial resolution in instrumentation is continually improving, so these types of measurements can likely be made in the future to determine how coherence lengths behave in this regime. Finally, fast radio bursts have recently been conjectured as a tool for estimating CGM cloud sized in \citet{mas-ribas2025}. By observing refractive scattering, clouds of $\lesssim$1 pc can be probed in proto-clusters at \z$\sim$2--3 or structures like high-velocity clouds at low redshift.

From the simulations in \citet{augustin2021}, the coherence length of the gas depends on the gas morphology (filaments or clumps). By determining these coherence lengths, we can better understand what the gas is physically doing (inflowing or outflowing) for MgII. It would be valuable to study the gas morphology of other ions (e.g., \CIV\ and \OVI) to determine if the different velocity components of the gas are in fact tracing different cloudlets. In \citet{rudie2019}, the \OVI\ profiles are matched across the 400 pc distance. Perhaps this ion can also be used to determine coherence lengths of the hotter phase of gas, while \HI\ can be used for the cooler phase. More investigations using simulations are needed to determine feasibility.

\subsection{Associated Galaxies}\label{sec:hosts}

Given the strong \HI\ and \CIV\ column densities of the two \CIV\ absorbers with associated \lya, it is likely they reside in the CGM of a foreground galaxy. Archival VLT/MUSE observations exist for the Sunburst Arc field (PID: 297.A-5012(A), PI: Aghanim), which we used to search for associated galaxies of the absorbers. MUSE has a field of view of 1$'$ $\times$ 1$'$, wavelength coverage from 4650--9300\AA, and a spectral resolution of $R$=2000-4000. Unfortunately, the redshift of the absorbers places all strong optical lines (starting with [\OII]$\lambda$$\lambda$3726,3728) outside of the MUSE wavelength range and strong UV lines (starting with \lya) not yet within this range. No galaxies at \z$\approx$2 were easily identifiable (S. Lopez 2024, private communication). It is possible the associated galaxies are not within the field of view, since it is centered on the foreground galaxy cluster and not on the Sunburst Arc. The hosts could also be lower luminosity dwarf galaxies in the field that are undetected in these observations (4449 seconds).

We also searched the \hst\ grism observations. In the infrared, the field is highly contaminated. The UVIS observations are less contaminated, but it is still difficult to identify an associated galaxy. Given that there are no obvious \z$\approx$2 galaxies, it is likely the associated galaxies are small and faint. We note there are several stars and large galaxies in the field that the host galaxies may also potentially be hidden under. 

Finally, we utilized ray-traced planes generated from the lensing model \citep{sharon2022} to identify galaxies close to the center of the field at \z$\approx$2. We followed these galaxies through several lower-redshift planes to \z=0. These objects are all small and faint in the \hst/ACS F814W image and cannot be identified in the MUSE observations to confirm their redshifts. With the current data of this field, we are unable to identify any associated galaxies for the \CIV\ absorbers with associated \lya. 

\section{Conclusion} \label{sec:concl}

In this paper, we have performed tomography measurements of IGM and CGM foreground absorbers and estimated their \HI\ masses using \hst/WFC3 UVIS G280 observations of the Sunburst Arc (\z$\approx$2.37). These tomography measurements are the first for pLLSs/LLSs at \z$\approx$2. The multiply-lensed LyC escape region of this galaxy provides a unique background source to probe the foreground gas with more than one sightline. Our main results are as follows:

\begin{enumerate}
\item We detect three foreground absorbers (two pLLSs and one LLS) at \z$\approx$2 with consistent \logNHI\ values across multiple sightlines in the LyC spectrum of the Sunburst Arc. Two of the absorbers are \CIV\ absorbers with associated \lya, and the other is a \lya\ absorber. 

\item The de-lensed lengths of the absorbers are all $\lesssim$2 kpc. Due to the unique geometry of the lensing system, a width measurement is also possible. They are upper limits and not constraining. 

\item Assuming the de-lensed length of the absorber is the diameter of a spherical cloud, we calculate \HI\ masses for the absorbers. The average is $\approx$$10^3$ \msun. 

\item The most likely locations for these absorbers are in the CGM for the \CIV\ absorbers with associated \lya\ and in the IGM for the \lya\ absorber. The high levels of \CIV\ when compared to the absorbers in the KODIAQ-Z survey indicate this gas is multiphase and collisionally-ionized. 

\item A search for the host galaxies of the \CIV\ absorbers with associated \lya\ did not yield any candidates. The existing data does not cover the optimal wavelength range for \z$\approx$2 redshift identifications. It is also possible the host galaxies lie outside of the field of view or are too faint to be identified in the observations. 
\end{enumerate}

The particular geometry of this system has allowed us to probe foreground pLLS and LLS gas clouds at small scales and in 2D for the first time at z$\approx$2. However, contamination in some sightlines did not allow us to put strict limits on the length scales. As such, these lengths are lower limits because the clouds could extend past the non-contaminated sightlines. It is also possible (and highly likely) that there is column density variations in the clouds to where different sightlines are not consistent with one another. This would also increase the cloud length scales. 

When comparing our results to other surveys at similar redshift, we see there are also consistent column densities for DLAs at 2-3 kpc \citep{bordoloi2022}. There is only one reported sub-kpc measure of a gas cloud from \citet{rudie2019}. At 400 pc separation, the total velocity spread of the cloud is similar, yet the individual components vary. The unfolding picture seems to point towards \HI\ tracing total structure lengths (coherence lengths), while metals may indicate either structure length or cloudlet sizes \citep{augustin2021}. Future observational work targeting sub-kpc length estimates are necessary to determine gas cloudlet structures. New gravitational arcs are being identified in \jwst\ observations of low redshift clusters previously observed with \hst. In the coming years, there may be many objects and opportunities to study sub-kpc cloud length scales. Observations of \HI\ and metal ions may also be able to together determine both cloud and cloudlet scales. Simulations investigating the structures that different ions trace will be critical in helping interpret these observations and guiding future survey design. 

\section*{Acknowledgments}

MAB thanks Marcel Neeleman, Michael Florian, Marc Rafelski, Gabriel Brammer, and Nor Pirzkal for their help with data reduction. MAB thanks Chris Howk and Nicolas Lehner for their useful discussions about the data. Support for this research was provided by NASA through grants HST-GO-15966 from the Space Telescope Science Institute, which is operated by the Association of Universities for Research in Astronomy (AURA), Incorporated, under NASA contract NAS5-26555. Some of the data used for this project were obtained from the KODIAQ project, which was funded through NASA ADAP grants NNX10AE84G and NNX16AF52G. This paper includes data gathered with the 6.5 m Magellan Telescopes located at Las Campanas Observatory, Chile. We thank the staff of Las Campanas for their dedicated service, which has made possible these observations. We thank the telescope allocation committees of the Carnegie Observatories, The University of Chicago, The University of Michigan, Massachusetts Institute of Technology, and Harvard University for supporting the MEGaSaURA project over several years of observing.

Facilities: \hst(WFC3), Magellan(MagE)

Software: {\small ASTROPY} \citep{astropy2013,astropy2018}, {\small BPASS} \citep{eldridge2017}, {\small DS9} \citep{ds92003}, {\small GRIZLI} \citep{brammer2019}, {\small LMFIT} \citep{lmfit2016}, {\small MATPLOTLIB} \citep{hunter2007}, {\small NUMPY} \citep{vanderwalt2011}, {\small PYNEB} \citep{luridiana2015}, {\small SPECUTILS} \citep{specutils2019}, {\small SCIPY} \citep{virtanen2020}, {\small XIDL} \citep{cooksey2023}

\section*{Data Availability}

All \hst\ data used in this paper can be found in MAST: \url{https://doi.org/10.17909/dezq-f603}. Data products are available upon request to the corresponding author. 

\bibliography{sunburst_lls}
\bibliographystyle{mnras}

\appendix

\section{BPASS Model Fits to the Knot Spectra}\label{sec:appa}

\begin{figure*}
   \centering
    \includegraphics[width=\textwidth]{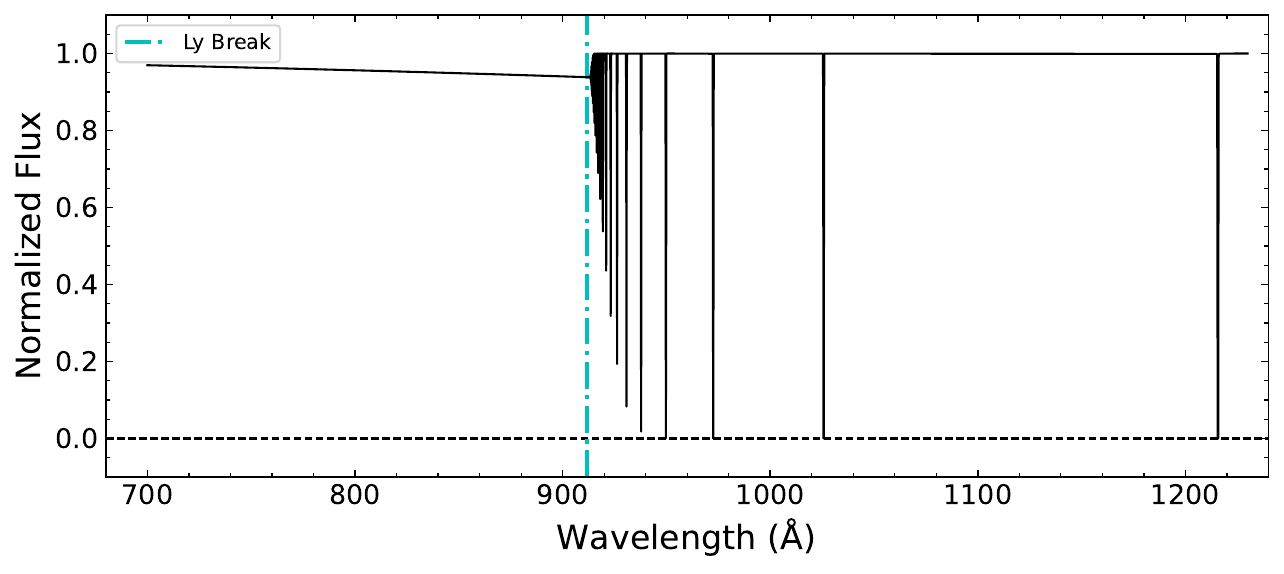}
    \caption{Model \HI\ absorber spectrum. The Lyman break of the absorber is designated with the dashed and dotted line, and the full series of Lyman lines are shown in absorption. This model is for an absorber with \logNHI=16. Larger column densities have increased absorption depth of the higher Lyman series lines, and the blueward flux from the break is also decreased. This model is interpolated to the UVIS wavelength grid and convolved with a 1D Gaussian filter with a FWHM=10\AA\ to mimic the UVIS line spread function prior to being added to the BPASS models.}
    \label{fig:llsmodel}
\end{figure*}

\begin{figure*}
   \centering
    \includegraphics[width=\textwidth]{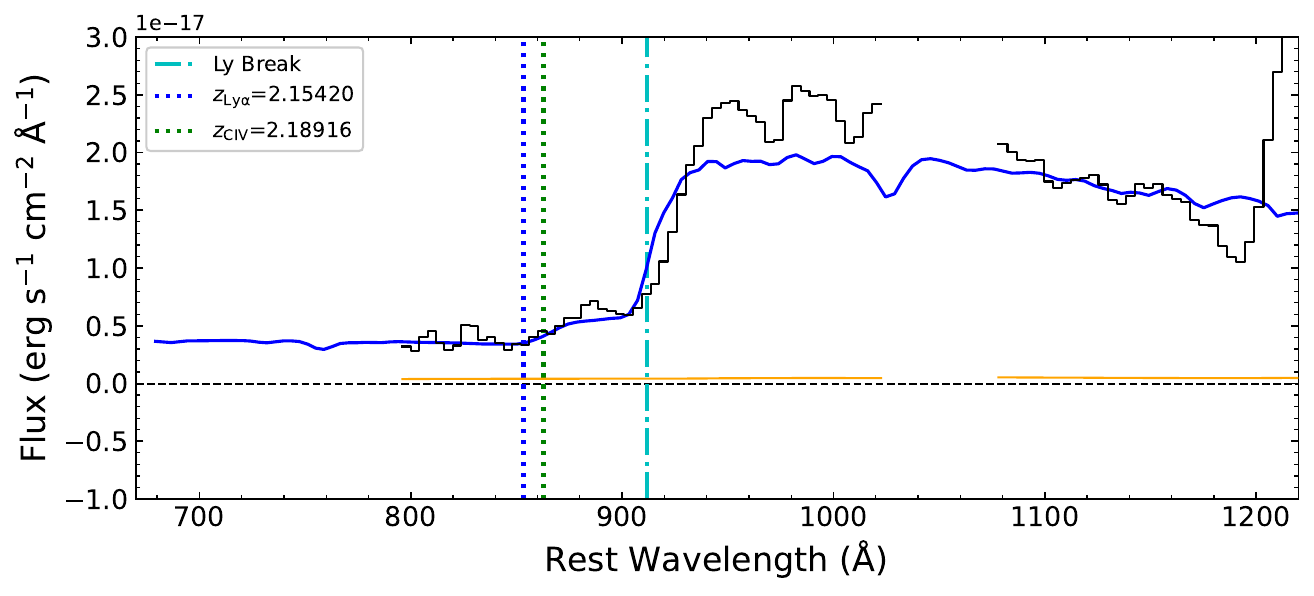}
    \caption{BPASS model fit to knot 2 spectrum. The galaxy Lyman break is denoted by the dashed and dotted line. The Lyman break of two intervening absorbers that appear in more than one knot spectrum are shown with the colored dotted lines. We do not fit the model to the data, but rather only add to the model the absorbers that are identified in the MagE spectrum if they improve the fit. Missing regions of flux denote masked contamination from another source. The model does not fit the redside data well because there is residual flux contamination that was not masked. This discrepancy has no bearing on the absorber fits, as the overall model fit to the data agrees well.}
    \label{fig:knot2}
\end{figure*}

\begin{figure*}
   \centering
    \includegraphics[width=\textwidth]{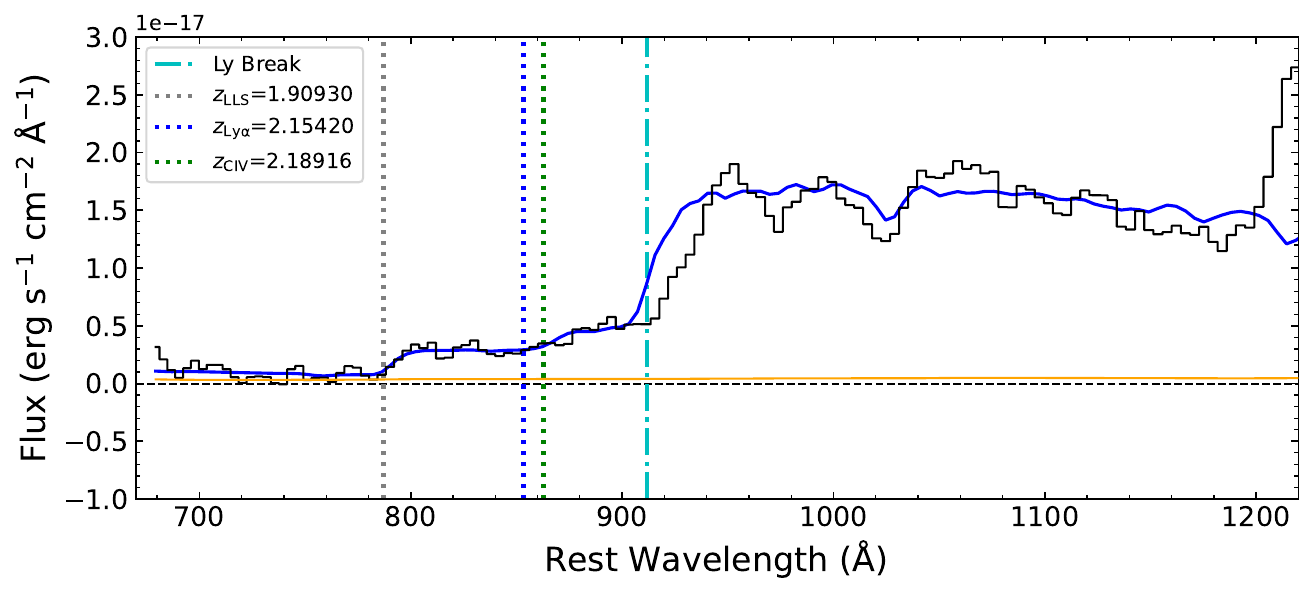}
    \caption{Same as Fig.~\ref{fig:knot2}, but for knot 5. The Lyman break of a singly-occurring absorber is shown with the gray dotted line.}
    \label{fig:knot5}
\end{figure*}

\begin{figure*}
   \centering
    \includegraphics[width=\textwidth]{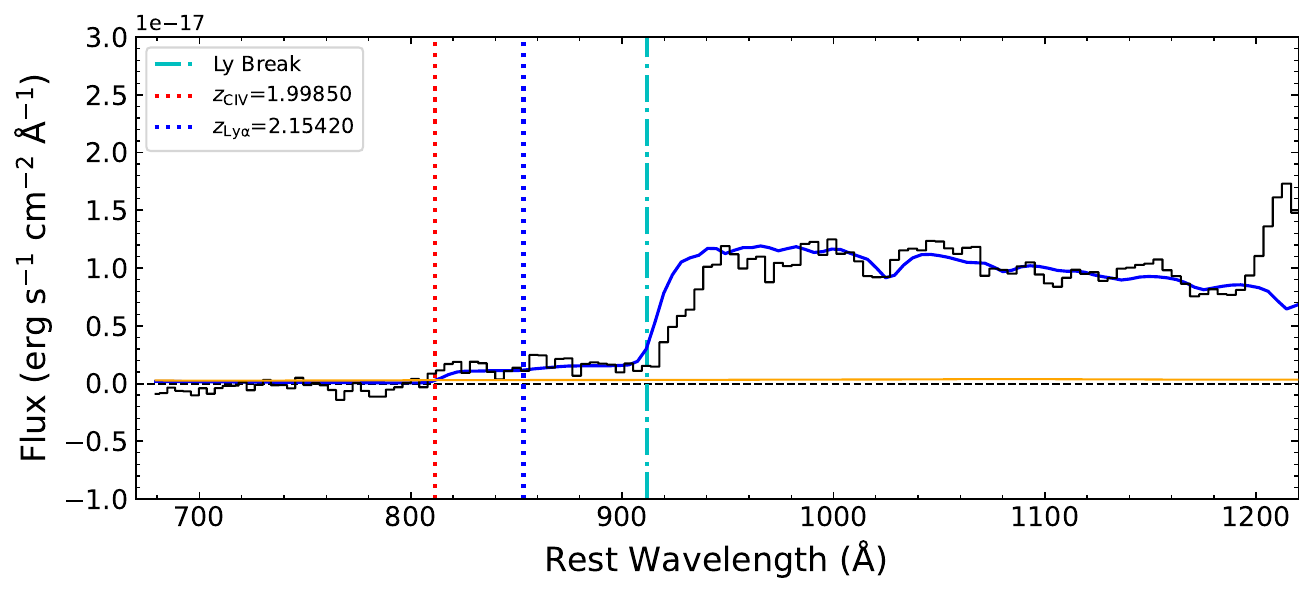}
    \caption{Same as Fig.~\ref{fig:knot2}, but for knot 6.}
    \label{fig:knot6}
\end{figure*}

\begin{figure*}
   \centering
    \includegraphics[width=\textwidth]{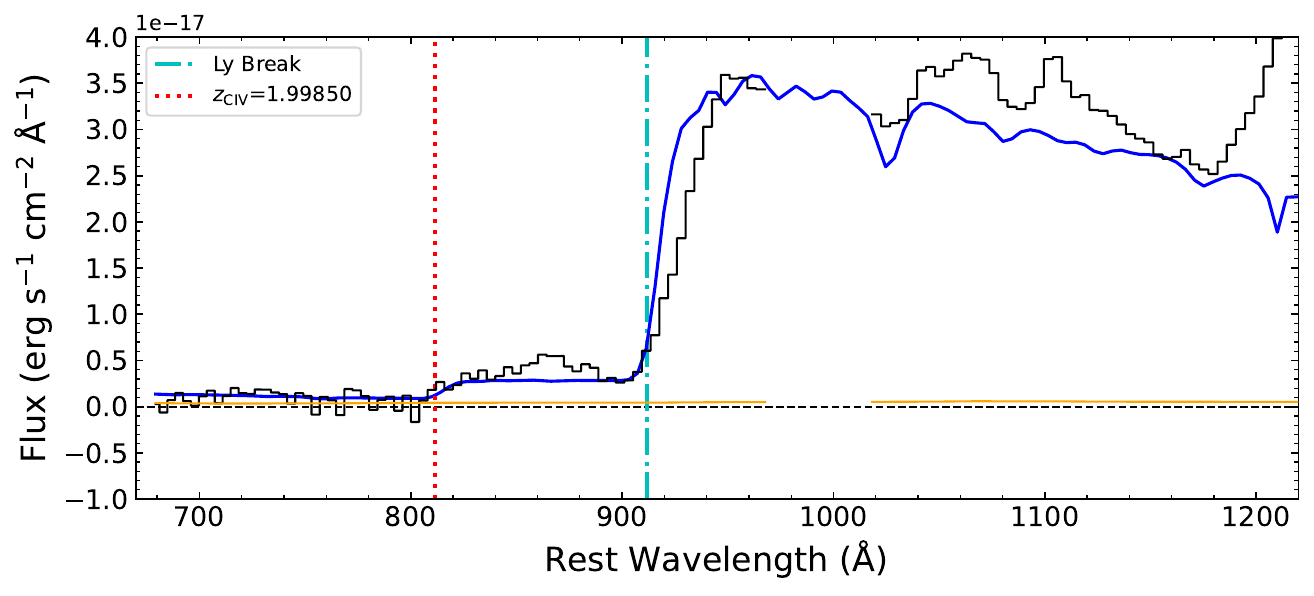}
    \caption{Same as Fig.~\ref{fig:knot2}, but for knot 10.}
    \label{fig:knot10}
\end{figure*}

\begin{figure*}
   \centering
    \includegraphics[width=\textwidth]{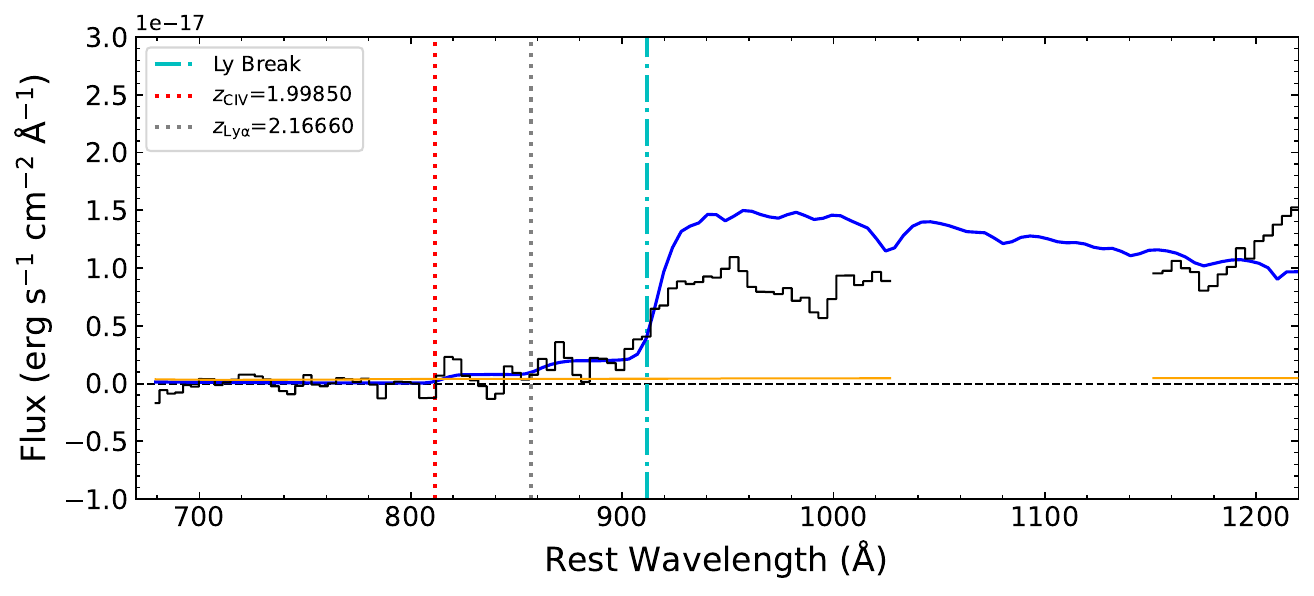}
    \caption{Same as Fig.~\ref{fig:knot2}, but for knot 11. The Lyman break of a singly-occurring absorber is shown with the gray dotted line. This knot has the lowest apparent magnitude, which is likely why the model does not fit the redside data well.}
    \label{fig:knot11}
\end{figure*}

\begin{figure*}
   \centering
    \includegraphics[width=\textwidth]{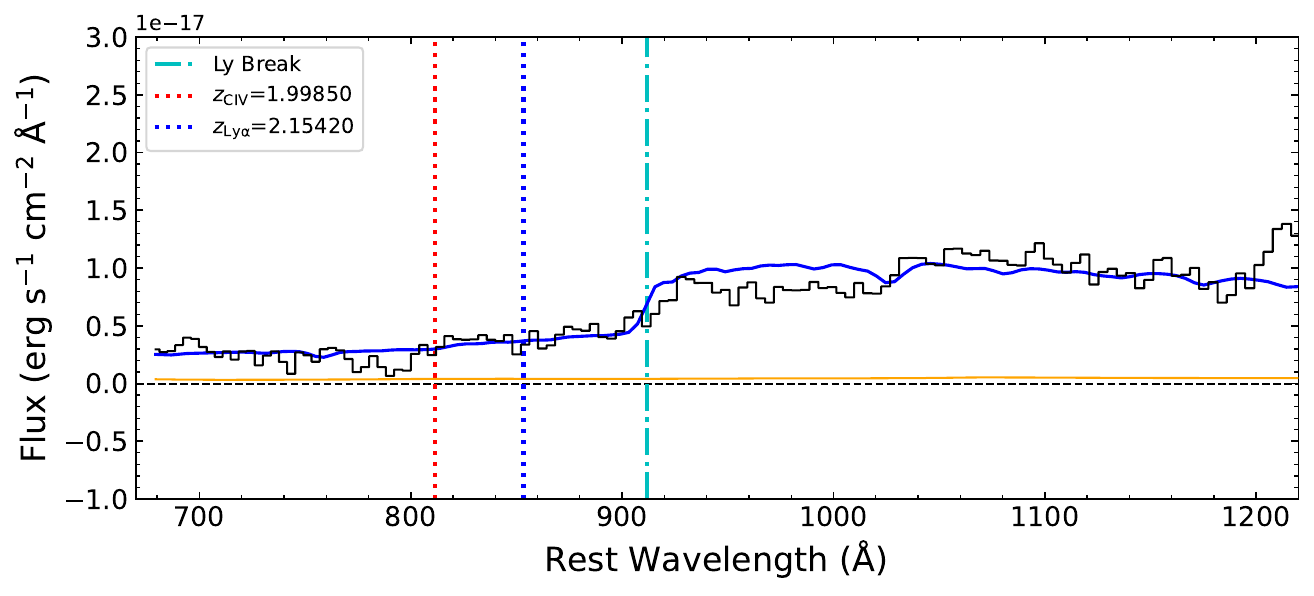}
    \caption{Same as Fig.~\ref{fig:knot2}, but for knot 12. The galaxy Lyman break did not need to be fit for this spectrum, implying it is dust-free.}
    \label{fig:knot12}
\end{figure*}

Here we present the model \HI\ absorber spectrum that was added to the BPASS models in Fig.~\ref{fig:llsmodel} and the other model fits of the knot spectra in Figures~\ref{fig:knot2}-\ref{fig:knot12}. Knots 1, 3, 8, and 9 are not included due to contamination that interferes with the absorber fitting. The spectrum for knot 7 was unable to be extracted.

\bsp	
\label{lastpage}
\end{document}